\documentclass[12pt, titlepage]{article}
\newcommand{\be}{\begin{equation}}
\newcommand{\ee}{\end{equation}}

\usepackage{epsf,rotate,psfig, latexsym, bbm}
\title{Causal Incompleteness: \\A New Perspective on Quantum Non-locality} \author{T.N.Palmer \\ 
ECMWF,\\ Shinfield Park,  Reading\\ RG2 9AX, UK\\tim.palmer@ecmwf.int} \begin{document} \maketitle

\begin{abstract}
The mathematical notion of incompleteness (eg of rational numbers, Turing-computable functions, and arithmetic proof) does not play a key role in conventional physics.  Here, a reformulation of the kinematics of quantum theory is attempted, based on an inherently granular and discontinuous state space, in which the quantum wavefunction is associated with a finite set of finite bit strings, and the unitary transformations of complex Hilbert space are reformulated as finite permutation and related operators incorporating complex and hyper-complex structure. Such a reformulation, consistent with Wheeler's `It from Bit' programme, provides the basis for a novel interpretation of the Bell theorem: that the experimental violation of the Bell inequalities reveals the inevitable incompleteness of the causal structure of physical theory. The kinematic reformulation of quantum theory so developed, provides a new perspective on the age-old dichotomy of free will versus determinism. 
\end{abstract}

\section{Introduction}
\label{sec:introduction}

It is often said that the most profound theorem of 20th Century mathematics concerns the incompleteness of arithmetic proof. The basis of G\"{o}del's theorem, via its use of the Cantor diagonal slash, is directly related to the incompleteness of the rational numbers and the Turing-computable functions. Thus incompleteness is rather fundamental in mathematics; and since mathematics completely underpins our scientific understanding of the physical world, one might ask whether incompleteness has any role to play in physical theory. The potential role of incompleteness in conventional physics is masked by its generic use of continuum equations. Here a novel interpretation of Bell's eponymous theorem of quantum physics is discussed, based on an attempt to recast the complex Hilbert space of quantum theory, using granular, discontinuous mathematics. In this interpretation, the experimental violation of the Bell inequalities reveals, not the type of non-local causality which mainstream physics regards as bizarre but unavoidable, but rather the inevitable incompleteness of the causal structure which may be inferred from such theory. 

The conventional proof of Bell's theorem presumes that the settings of measuring instruments can be treated as free variables. That is, for a given entangled particle pair, it is assumed that the causal consequences of choosing alternate instrument settings on measurement outcomes are well defined.  Bell(1993) himself realised that this issue was not metaphysically clear cut, since the world is given to us once only: `we cannot repeat an experiment changing just one variable; the hands of the clock will have moved and the moons of Jupiter'. However, for Bell, the existence or otherwise of free variables was something to be inferred from the mathematical structure of physical theory, rather than from metaphysical analysis. Hence, for example, if 
\be
\label{eq:det}
\dot{\mathbf{X}}=\mathbf{F}[\mathbf{X}]
\ee
denotes a conventional continuum evolution equation, such as occurs in standard quantum theory, electromagnetism, general relativity as so on, then (\ref{eq:det}) determines not only how a given initial state vector $\mathbf{X}(0)$ evolves at future times $t>0$, but also the causal consequences at $t>0$, of a hypothetical perturbation $\delta X$, for example to one of $\mathbf{X}(0)$'s components. In this respect, most physicists (the author included) would agree with Bell(1993)that his eponymous theorem `is primarily an analysis of certain kinds of physical theory', metaphysical concerns notwithstanding. 

On the other hand, the conventional non-locally causal interpretation of Bell's theorem, that the influence of some freely-chosen remote instrument setting can propagate through space at superluminal speed, remains as bizarre and incomprehensible today as when it was first propounded. Although there have been other proposals to understand the Bell theorem, such as backwards-in-time causality (Price, 1996), one might ask whether there exist classes of theory, formulated using less conventional mathematical structures to that of (\ref{eq:det}) above, for which the existence of an unrestricted set of causal consequences between alternate detector orientations and measurement outcomes, cannot be assumed.  
The purpose of this paper is to analyse `certain kinds of physical theory` for which the freedom to perturb mathematically the values of certain key variables, is determined by the underlying mathematical structure of the theory's state space. Of particular relevance here will be systems whose state space is generically granular and discontinuous (that is, state space cannot be `continued' by Cauchy-sequence methods). 

In Section \ref{sec:toy} an idealised model is outlined as the basis for a discussion of the standard EPR-Bohm-Bell experiment, in which the role of causal incompleteness is made explicit. Here we make use of an elementary property of the cosine function (albeit one that the author is unaware of having been used before in physics): if $0<\cos \theta< 1$ is rational, $\theta/\pi$ does not have a finite binary expansion: see Appendix 1 for a simple proof of this. 

In Section \ref{sec:q} is discussed a potential reformulation, granular and discontinuous, of the kinematics of conventional quantum theory. In this reformulation, the wavefunction becomes a set of finite bit strings and unitary transforms become permutation and related operators with inherent complex and hyper-complex structure. This kinematic reformulation may be of direct interest in quantum information theory as a novel attempt to define physical reality consistent with J.A. Wheeler's `It from Bit' programme (Wheeler, 1994). The analysis of causal incompleteness and the Bell theorem is discussed in section \ref{sec:toy} is relation to this reformulation. 

Some discussion of the notion of causal incompleteness in the context of the age-old cognitive dichotomy of free-will versus determinism, is given in Section \ref{sec:meta}. It is suggested that the type of mathematically-developed causal incompleteness discussed in sections \ref{sec:toy} and \ref{sec:q}, may present a new perspective on the age-old dichotomy of free will and determinism (Kane, 2002). 

\section{Quantum Entanglement and Causal Incompleteness: A Simplified Model}
\label{sec:toy}
The main goal of this paper is to outline a possible reformulation of the kinematics of quantum theory, in which the incompleteness of causal structure can be made explicit. Before doing so, a simplified deterministic model is developed, consistent with two-particle quantum entanglement statistics, which illustrates the potential role of causal incompleteness in the interpretation of the Bell theorem. Linkage between this simplified model and the proposed reformulation of the kinematics of quantum theory is developed in the next section. 

\subsection{Preliminaries}

Let $\mathcal{N}_0$ denote a dyadic rational, ie a member of the set $\mathbbm{Q}_2$ of numbers with finite binary expansion. For example, let the binary expansion of $\mathcal{N}_0$ agree with the first $2^N$ bits of the binary Champernowne number $=.11011100101\ldots$, formed by concatenating the natural numbers $1,2,3,4\ldots$ in binary representation. With $c_n$ denoting the $n$th bit of $\mathcal{N}_0$, let $a_n=2c_n-1 \in \{1,-1\}$. Then, with  $\{\lambda_n=\frac{n-1}{2^N}\pi: 1\le n \le 2^N\}$ denoting a finite subset of the unit semi-circle $0 \le \lambda \le \pi$, define 
\be
\label{eq:s}
S_0(\lambda_n)=a_n \ \ \ \ \ \ S_0(\lambda_n+\pi)=-a_n.
\ee
For sufficiently large $N$, $S_0$ is defined on arbitrarily-dense subsets of the unit circle.

More generally, let $\mathcal{N}_{\theta}$ be the dyadic rational obtained by flipping ($0 \rightarrow 1, 1 \rightarrow 0$) every $1/\sin^2 \frac{\theta}{2}$th bit of $\mathcal{N}_0$, where $\cos \theta \in \mathbbm{Q}_2$. Hence, for example, with $\mathcal{N}_0$ based on the Champernowne number, then
\be
\mathcal{N}_{\pi/2}=.10001001111\ldots \ \ \ \mathcal{N}_{\pi}=.00100011010\ldots
\ee 
Let $c_n(\theta)$ denote the corresponding bits of $\mathcal{N}_{\theta}$, $a_n(\theta)=2c_n(\theta)-1$, and define
\be
\label{eq:s2}
S_{\theta}(\lambda_n)=a_n(\theta) \ \ \ \ \ \ S_{\theta}(\lambda_n+\pi)=-a_n(\theta)
\ee

Since $\mathcal{N}_0$ is based on a normal number (Hardy and Wright, 1979), for large enough $N$ the values of either $S_0$ or $S_{\theta}(\lambda)$, sampled over $S_0$-defined points in any subset of the unit circle, comprise equal numbers of +1's and -1's. Moreover, by construction, the corresponding sample coefficient of correlation 

\be
\label{eq:correal}
C(\theta)=\langle \ S_0(\lambda)S_{\theta}(\lambda+\pi)\rangle =-\cos\theta
\ee

\subsection{A Specific Reality}
\label{sec:reality}

We now add some interpretational baggage. Imagine two experimenters, each with Stern-Gerlach detectors, measuring the spin of entangled particle pairs in a standard EPR-Bohm-Bell experiment. The orientation of experimenter 1's detector is defined to be the $z$-axis; the orientation of experimenter 2's detector is at angle $\theta$ to the $z$ axis (corresponding to a rotation about the particle beam axis). Let $S_0(\lambda)$ and $S_{\theta}(\lambda+\pi)$ determine the measurement outcomes for a given entangled particle pair labelled by $\lambda$. (Nb if $\mathcal{M}_0=\mathcal{N}_{\theta}$ had been used in place of $\mathcal{N}_0$ as the generating base-2 normal number, then experimenter 1's spin measurement outcomes would be determined by $\mathcal{M}_{\theta}$.)

Consider a $\theta(t)$ in which the experimenters' detectors have relative orientation $\theta=\theta_A$ when $t_1<t<t_2$ and $\theta=\theta_B$ when $t_3<t<t_4$. We have 
\begin{eqnarray}
\label{eq:corr}
C(\theta_A)&=&\langle \ S_0(\lambda)S_{\theta_A}(\lambda+\pi)\ \rangle=-\cos\theta_A \nonumber \\
C(\theta_B)&=&\langle \ S_0(\lambda)S_{\theta_B}(\lambda+\pi)\ \rangle=-\cos \theta_B
\end{eqnarray}  
consistent with quantum experimentation. The values $S_0(\lambda)$ and $S_{\theta}(\lambda)$ associated with such a $\theta(t)$ are referred to as a `specific reality'. The space $\mathcal{U}(\mathcal{N}, \theta(t))$ of such `specific realities' can be generated, as far as this idealised model is concerned, by varying over continguous length-$2^N$ segments $\mathcal{N}$ of the Champernowne number, and timeseries $\theta(t)$ where, for all $t$, $\cos\theta(t) \in \mathbbm{Q}_2$. 

\subsection{Causal Extension of the Specific Reality}

In the introduction, it was noted that (\ref{eq:det}) determined not only the evolution $\mathbf{X}(t)$ from some specific initial state, but also the causal effect on $\mathbf{X}(t)$ of a perturbation $\delta \mathbf{X}$ to that initial state. If the system is ergodic, then this causal effect is, in principle, determined from knowledge of the evolution $\mathbf{X}(t)$. 

In keeping with our intuition that the experimenters are free to choose individually the orientations of their detectors, it can similarly be asked whether the functions which describe the `specific reality' above, $S_0(\lambda)$ and $S_{\theta}(\lambda)$, also provide the information required to describe the causal effect on measurement outcome, of hypothetical perturbations $\delta\theta_1$ and $\delta\theta_2$ to experimenter 1 and 2's actual detector orientations.   

To this end, consider the functions
\begin{eqnarray}
\label{eq:cf}
Sp_1(\delta\theta_1, \lambda)&=&\ \ S_0(\lambda-\delta\theta_1)\nonumber\\
Sp_2(\delta\theta_2, \lambda)&=&-S_{\theta}(\lambda-\delta\theta_2)
\end{eqnarray}
written in conventional local hidden-variable form.  When $\delta\theta_1=\delta \theta_2=0$, $Sp_1$ and $Sp_2$ describe the specific reality above, ie 
\begin{eqnarray}
Sp_1(0, \lambda)&=&\ \ S_0(\lambda)\nonumber\\
Sp_2(0, \lambda)&=&-S_{\theta}(\lambda)=S_{\theta}(\lambda+\pi)
\end{eqnarray}
Hence, assume that $Sp_1(\delta\theta_1, \lambda)$ determines the spin value of one particle of an entangled particle pair labelled by $\lambda$ under a hypothetical perturbation $\delta\theta_1$ to the orientation of experimenter 1's detector. Similarly, let $Sp_2(\delta\theta_2, \lambda)$ determine the spin value of the other entangled particle under a hypothetical perturbation $\delta\theta_2$ to the orientation of experimenter 2's detector. That is, we can think of (\ref{eq:cf}) as defining a pair of lists which give the causal consequences on measurement outcome of hypothetical perturbations to detector orientations. Since, for $N \rightarrow \infty$, $Sp_1$ and $Sp_2$ are defined on uniformly dense subsets of the circle, the functions $Sp_1$ and $Sp_2$ would, for sufficiently large $N$, appear to accommodate any hypothetical alternate choices of orientation experimenter 1 and 2 would care to make. 

From the construction of $S_0$ and $S_{\theta}$ above,  a necessary condition that $Sp_1$ and $Sp_2$ be defined is that both $(\lambda-\delta\theta_1)/\pi$ and $(\lambda-\delta\theta_2/)\pi$ belong to $\mathbbm{Q}_2$. On the other hand, in order that $Sp_1$ and $Sp_2$ describe one of the specific realities of section \ref{sec:reality}, the cosine of the hypothetical relative orientation $\Delta \theta = \theta+(\delta\theta_2-\delta\theta_1)$ must be dyadic rational. There are certainly occasions where (\ref{eq:cf}) returns the correct quantum correlations when $\delta \theta_1 \ne 0$. For example, for all $\delta\theta_1=\delta\theta_2=\delta\theta'$, $\cos \Delta \theta = \cos \theta$ which by construction belongs to $\mathbbm{Q}_2$. In this situation, the hypothetical correlation $\langle \ Sp_1(\delta \theta', \lambda)Sp_2(\delta \theta', \lambda)\ \rangle$ is invariant under hypothetical identical perturbations $\delta \theta'$ to the orientations of both detectors. On the other hand, in general, it cannot be assumed that $\cos \Delta \theta \in \mathbbm{Q}_2$, the implications of which are discussed in the next section. 

\subsection{Bell's Theorem and Causal Incompleteness}

As written, (\ref{eq:cf}) is local in terms of the hypothetical detector perturbations; $Sp_1$ does not depend on $\delta\theta_2$, and $Sp_2$ does not depend on $\delta\theta_1$. Does this imply that correlation statistics derived from (\ref{eq:cf}) must satisfy a Bell inequality? In order to derive a Bell inequality from the statistics of the two experiments with $\theta=\theta_A$ and $\theta=\theta_B$, we need to assume what, following EPR, are usually called `Reality Conditions'. For the first experiment where $\theta=\theta_A$, the relevant Reality Condition states that if a hypothetical perturbation $\delta\theta_1=\theta_A$ to the orientation of experimenter 1's detector were to have aligned experimenter 1's detector with experimenter 2's detector, then experimenter 1 would have measured exactly the opposite of experimenter 2, ie 
\be
\label{eq:reality}
Sp_1(\theta_A, \lambda)=-Sp_2(0, \lambda)
\ee
For the second experiment ($\theta=\theta_B$), the Reality Condition similarly requires
\be
\label{eq:reality2}
Sp_1(\theta_B, \lambda)=-Sp_2(0, \lambda)
\ee
The anti-correlations expressed in (\ref{eq:reality}) and (\ref{eq:reality2}) should be contrasted with the anti-correlations, which by the definitions of $S_0(\lambda)$ and $S_{\theta}(\lambda+\pi)$, are guaranteed in the subset of occasions when $\theta=0$ (ie both detectors aligned with the $z$ axis) within the specific reality defined by $\theta=\theta(t)$. The difference between these two situations is exactly equivalent to the difference between the counterfactual and regularity definitions of causality (Menzies, 2001) as first enunciated by the philosopher David Hume. If (\ref{eq:reality}) and (\ref{eq:reality2}) are assumed, then the Bell inequalities follow by standard text-book analysis (eg Rae, 1992). 
 
However, in the present case, it must be asked whether (\ref{eq:reality}) and (\ref{eq:reality2}) are consistent with the global constraint $\cos \Delta \theta \in \mathbbm{Q}_2$. Consider (\ref{eq:reality}) in particular. Putting $\delta\theta_1=\theta_A$, $\delta\theta_2=0$ in (\ref{eq:cf}), we have 
\begin{eqnarray}
\label{eq:crunch}
Sp_1(\theta_A, \lambda)&=&S_0(\lambda-\theta_A)\nonumber\\
Sp_2(0, \lambda)&=&-S_{\theta_A}(\lambda)
\end{eqnarray}
But now the result in Appendix A becomes relevant. Since $\cos\theta_A$ is required to be dyadic rational, $\theta_A$ cannot be a dyadic rational fraction of $\pi$. Now, from (\ref{eq:crunch}), $Sp_2$ is only defined if $\lambda$ is a dyadic rational fraction of $\pi$. Hence, if $\lambda$ is a dyadic rational fraction of $\pi$, and $\theta_A$ not, then $(\lambda-\theta_A)/\pi\ \notin \mathbbm{Q}_2$. Hence, for given $\lambda$, ie entangled particle pair, $Sp_1(\theta_A, \lambda)$ and $Sp_2(0, \lambda)$ are not simultaneously defined (reminiscent of the Principle of Complementarity), no matter how large is $N$. Alternatively, $Sp_1(\theta_A, \lambda)$ and $Sp_2(0,\lambda)$ are not contained in the lists of causal relations defined by (\ref{eq:cf}), even, as $N \rightarrow \infty$, this list becomes infinitely long. A similar conclusion holds for (\ref{eq:reality2}). Hence, we cannot derive a Bell inequality from the correlation statistics associated with (\ref{eq:cf}). 

Is it not instead possible to derive a Bell inequality using in (\ref{eq:reality}) a hypothetical perturbation $\delta\theta$ which is a good dyadic rational approximation $\theta'_A$ to $\theta_A$? Since, the hidden-variable model (\ref{eq:cf}) is generically discontinuous, it is not possible. That is to say, if we consider a Cauchy sequence $\{\theta'_A, \theta''_A, \theta'''_A,\ldots\}$ of dyadic rational perturbations which converge on $\theta_A$, the corresponding sequence $Sp_1(\theta'_A, \lambda), Sp_1(\theta''_A, \lambda), Sp_1(\theta'''_A, \lambda), \ldots$ will not converge to some well-defined value $Sp_1(\theta_A, \lambda)$. 

Hence although a hidden-variable model has been defined, whose support is as dense as we like on the cirle, generating in the limit $N \rightarrow \infty$ an infinite set of causal relationships between measurement outcome and hypothetical perturbation to detector orientation, the causal structure is not sufficiently comprehensive to be able to derive a Bell inequality. Could quantum theory be recast in such a form as to be able to exploit this result?

\section{It from Bit - Towards A Theory of Quantum Beables}
\label{sec:q}

In this section an attempt to reformulate the kinematics of quantum theory as a generically granular and discontinuous theory is outlined, in order to exploit the notion of causal incompleteness, as discussed above. In this reformulation, the quantum wavefunction $|\psi\rangle$ is a set of encoded bit strings, and the unitary transformations are self-similar permutation and related operators with complex and hyper-complex structure. As such, the wavefunction is literally identified with `information', consistent with J.A.Wheeler's (1994) `It from Bit' aphorism, and Bell's notion of beables. There is no additional collapse hypothesis. The reformulation renders the state space of the wavefunction as inherently granular and discontinuous, yielding a mathematical structure from which the notion of incompleteness, discussed above, is manifest. 

In this reformulation, the wavefunction of an elementary 2-level system will be identified with the single bit string   
\be
\label{eq:s3}
\mathcal{S}=\{a_1, a_2, a_3 \ldots, a_{2^N}\}
\ee
where $a_i \in \{1,-1\}$ and $N \gg 1$ denotes the number of such 2-state systems in the universe. The wavefunction of the universe as a whole is given by $N$ such bit strings; equivalently, a single rational $\mathcal{R}_U$ defined from a length-$2^N$ string comprising base-$2^N$ digits. The entanglement structure of the universe is defined by non-zero coefficients of correlation between individual bit strings; equivalently, in unequal frequencies of occurrence of the digits in the base-$2^N$ expansion of $\mathcal{R}_U$. Following Bohm (1980), this non-normal number structure could be referred to as the `implicate order', whereas the apparently random sequence of bits in any one string could be referred to as the `explicate order'.  In the following, some emphasis is placed on ensuring that the proposed reformulation can correctly account for the `vastness' of Hilbert space. 

\subsection{Permutation Operators with Complex and Hyper-Complex Structure}
One of the key features of quantum theory is that its state space is complex. Here we introduce complex structure through permutation operators $\mathbf{E}$, acting on bit strings $\mathcal{S}$, which satisfy
 \be
\label{eq:sqrt}
\mathbf{E}^2(\mathcal{S})=-\mathcal{S}=\{-a_1, -a_2, -a_3 \ldots, -a_{2^N}\}.
\ee
We start by defining such complex structures starting with the simplest $N=1$ `universe', and build up complexity for higher $N$.  
\subsubsection{N=1}
With $\mathcal{S}=\{a_1,a_2\}$, define 
\be
\label{eq:i}
\mathbf{i} (\mathcal{S})  =\{a_2, -a_1\}
\ee
so that $\mathbf{E}=\mathbf{i}$ satisfies (\ref{eq:sqrt}). It is convenient to rewrite (\ref{eq:i}) as $\mathbf{i}(\mathcal{S}) =\{a_1, a_2\}i$ interpreting $\{a_1, a_2\}$ as a row vector, and 
\be
\label{eq:im}
i=\left( \begin{array}{cc}
0&1\\
-1&0 
\end{array} \right)
\ee
The coefficient of correlation between $\mathcal{S}$ and $\mathbf{i}(\mathcal{S})$ is equal to zero. 

\subsubsection{N=2}
With $\mathcal{S}=\{a_1,a_2, a_3, a_4\}$, define
\begin{eqnarray}
\mathbf{e}_1(\mathcal{S})&=&\{-a_3, -a_4, a_1, a_2\}\nonumber\\
\mathbf{e}_2(\mathcal{S})&=&\{-a_4, a_3, -a_2, a_1\}\nonumber\\
\mathbf{e}_3(\mathcal{S})&=&\{-a_2, a_1, a_4, -a_3\}
\end{eqnarray}
In matrix notation, this can be written, for $j=1,2,3$, as
\be
\mathbf{e}_j (\mathcal{S}) =\{a_1, a_2, a_3, a_4\}e_j
 \ee
where $e_j$ are $4\times 4$ matrices in block $2 \times 2$ form
\be
\label{eq:m2}
e_1=\left(
\begin{array}{cc}
0&1\\
-1&0
\end{array}\right),\ 
e_2=\left(
\begin{array}{cc}
0&i\\
i&0
\end{array}\right),\ 
e_3=\left(
\begin{array}{cc}
i&0\\
0&-i
\end{array}\right) 
\ee
These matrices satisfy the laws of quaternionic multiplication, ie
\be
e_j e_j = e_1 e_2 e_3 = -\mathrm{Id}
\ee
and, hence, in particular, each $\mathbf{E}=\mathbf{e}_j$ satisfies (\ref{eq:sqrt}). Note that $e_1$ has the same block form as $i$ in (\ref{eq:im}). With $\mathbf{e}_0$ equal to the identity, the coefficient of correlation between any pair of sequences $(\mathbf{e}_j(\mathcal{S}), \mathbf{e}_k(\mathcal{S}))$, with $0 \le j,k \le 3$, is equal to zero. 

\subsubsection{N=3}
Based on the quaternions above, we can, by self-similarity, construct 7 independent square-root-of-minus-one permutation operators acting on 8-element sequences $\mathcal{S}=\{a_1, a_2, \ldots a_8\}$, ie for $j=1,2 \ldots 8$,
\be
\mathbf{E}_j(\mathcal{S})=\{a_1, a_2, \ldots, a_8\} E_j
\ee
where $E_j$ are $8\times8$ matrices in block $2 \times 2$ form
\begin{eqnarray}
\label{eq:m3}
E_1&=&\left(
\begin{array}{cc}
0&\ \ 1\\
-1&\ 0
\end{array}\right),\  \nonumber \\
E_2=\left(
\begin{array}{cc}
0&e_1\\
e_1&0
\end{array}\right), \ 
E_3&=&\left(
\begin{array}{cc}
0&\ e_2\\
\ e_2&0
\end{array}\right),\ 
E_4=\left(
\begin{array}{cc}
0&e_3\\
e_3&0
\end{array}\right) \nonumber\\ 
E_5=\left(
\begin{array}{cc}
e_1&0\\
0&-e_1
\end{array}\right), \ 
E_6&=&\left(
\begin{array}{cc}
e_2&0\\
0&-e_2
\end{array}\right),\ 
E_7=\left(
\begin{array}{cc}
e_3&0\\
0&-e_3
\end{array}\right)
\end{eqnarray}  
Note that $E_1$ has the same block form as $i$ in (\ref{eq:im}), and each $E_j$, $j>1$ belongs to one of the pure imaginary quaternion triples $\{E_1, E_2, E_5\}$, $\{E_1,E_3,E_6\}$ and $\{E_1, E_4, E_7\}$. With $\mathbf{E}_0$ equal to the identity, the coefficient of correlation between any pair of sequences $(\mathbf{E}_j(\mathcal{S}), \mathbf{E}_k(\mathcal{S})$, $0\le j,k \le 7$ is equal to zero.

\subsubsection{Arbitrary $N$}
The construction above can be continued, by self similarity, to $N=4, 5 \ldots$. For arbitary $N$ we have $2^N-1$ square-root-of-minus-one permutation matrices $E_1, E_2 \ldots, {E}_{2^N-1}$, each of which can be written as a $2 \times 2$ block matrix, with blocks representing $2^{N-1}\times 2^{N-1}$ matrices. The square roots are orthogonal to one another, and to the identity $E_0$ in the sense that the coefficient of correlation between any pair $(\mathbf{E}_j(\mathcal{S}), \mathbf{E}_k(\mathcal{S})$, $0\le j,k \le 2^N-1$, is equal to zero. 
 
By self-similarity, each of the $M<N$th sets of square roots of minus one, eg (\ref{eq:m2}) and (\ref{eq:m3}), are embedded in this larger $N$th set. 

Focus now on the matrix $E_1 \equiv E$ which has the special block form 
\be
\label{eq:E}
E=\left( \begin{array}{cc}
0&1\\
-1&0
\end{array} \right)
\ee
similar to (\ref{eq:im}), but where `$0$' and `$1$' denote the $2^{N-1} \times 2^{N-1}$ zero and identity matrices, respectively. $E$ has a square root which can be written as the $4\times4$ block matrix
\be
E^{1/2}=
\left( \begin{array}{cccc}
0&1&0&0\\
0&0&1&0\\
0&0&0&1\\
-1&0&0&0
\end{array} \right)
\ee
where `$0$' and `$1$' now denote the $2^{N-2} \times 2^{N-2}$ zero and identity matrices, respectively. In turn, $E^{1/2}$ has a square root which can be written as the $8\times8$ block matrix
\be
E^{1/4}=
\left( \begin{array}{cccccccc}
0&1&0&0&0&0&0&0\\
0&0&1&0&0&0&0&0\\
0&0&0&1&0&0&0&0\\
0&0&0&0&1&0&0&0\\
0&0&0&0&0&1&0&0\\
0&0&0&0&0&0&1&0\\
0&0&0&0&0&0&0&1\\
-1&0&0&0&0&0&0&0
\end{array} \right)
\ee
where `$0$' and `$1$' now denote the $2^{N-3} \times 2^{N-3}$ zero and identity matrices, respectively. This procedure can be continued until we reach the $2^N$th root given by the $2^N \times 2^N$ matrix
\be
\label{eq:root}
E^{1/2^N}=\left(
\begin{array}{cccccc}
0&1&\ &\ &\ldots&0\\
0&0&1 &\ &\  &0\\
0&0&0&1&\  &0\\
\ &\ &\ &\ &\ddots&\ \\
0&0&0&0&\ldots&1\\
-1&0&0&0&\ldots&0
\end{array} \right)
\ee
where `$0$' and `$1$' are scalars. Clearly $E^{1/2^N}$ is a $2^{N+2}$th root of unity and therefore generates a cyclic group of order $2^{N+2}$, a finite sub-group of $U(1)$. Applied to a bit string $\mathcal{S}=\{a_1, a_2, \ldots, a_{2^N}\}$, $E^{1/2^N}(\mathcal{S})=\{-a_{2^N}, a_1, a_2, \ldots, a_{2^N-1}\}$; that is, $E^{1/2^N}$ brings to the front, the (negation of the) trailing bit of $\mathcal{S}$, cf the discussion in section \ref{sec:meta}. 

\subsection{Towards A Granular Reformulation of Complex Hilbert Space}

Here the results above are applied to a possible kinematic reformulation of quantum theory. Some preliminary ideas on such a reformulation were first presented in Palmer (2003). A more complete exposition will appear elsewhere. 

\subsubsection{1 Qubit}
\label{sec:1qubit}

The general 1-qubit state in quantum theory is given, for example, by
\be
\label{eq:qubit}
|\psi\rangle=p_0|\uparrow\rangle+p_1|\downarrow\rangle)
\ee 
where $p_0, p_1 \in \mathbbm{C}$ satisfy $|p_0|^2+|p_1|^2=1$. The Hilbert space of such a qubit is therefore three dimensional, including the phase degree of freedom
\be
\label{eq:gphase}
|\psi\rangle \mapsto e^{i\phi}|\psi\rangle
\ee
which in quantum theory is viewed as `irrelevant' since the value $\phi$ does not affect the statistics of measurement outcomes. In the proposed reformulation proposed here, $|\psi\rangle \mapsto \mathcal{S}$, see (\ref{eq:s3}). The elements of $\mathcal{S}$ can be associated with what, operationally, are measurement outcomes, but which, following Bell, might be better called `beables'; $\mathcal{S}$ can therefore be thought of as a time series of such beables, whose first element is the beable corresponding to `now'. 

To account for the three degrees of freedom of complex Hilbert space, recall from (\ref{eq:m2}) that, in an $N$ qubit universe, any of the $2^N-2$ square roots of minus one, ${E}_j \ j>1$, is automatically part of the (pure imaginary) quaternionic triple $\{E, E_j, {E}_{j+2^{N-1}-1}\}$ if $j \le 2^{N-1}$, or the triple $\{E, E_{j-2^{N-1}+1}, E_j\}$ if $j>2^{N-1}$, where $E$ is given by (\ref{eq:E}). We build the reformulation of the qubit state (\ref{eq:qubit}) around one such quaternion triple, written in the generic form $\{E, E_a, E_b\}$. Then, 
\begin{eqnarray}
|\uparrow\rangle &\mapsto& \{1,1,\ldots,1\}\nonumber\\
|\uparrow\rangle+|\downarrow\rangle &\mapsto& \mathbf{E}_a ( \{1,1,\ldots,1\})\nonumber\\
|\downarrow\rangle &\mapsto& \{0,0,\ldots,0\}.
\end{eqnarray}

Let us start with the degree of freedom associated with the phase transformation (\ref{eq:gphase}), here reformulated as $\mathcal{S} \mapsto \mathbf{E}^{2\phi/\pi}(\mathcal{S})$. Consistent with the invariance of the qubit wavefunction under a global phase transformation in standard quantum theory, here a qubit is regarded as an equivalence class of bit strings, where two bit strings $\mathcal{S},\ \mathcal{S}'$ in the class are related by $\mathcal{S}'=\mathbf{E}^{\alpha}(\mathcal{S})$, $\alpha \in \mathbbm{Q}_2$. 

To reformulate the `nontrivial' unitary transformations associated with the remaining two degrees of freedom, those which transform one qubit state into a physically-inequivalent qubit state, we define the notion of addition of sequences eg as in 
\be
\label{eq:superp}
\mathcal{S}=(\cos\theta\  \mathbf{E}_a + \sin\theta\  \mathbf{E}_b) (\{1,1,\ldots,1\})
\ee
as follows. If $\cos\theta \in \mathbbm{Q}_2$, then the $n$th element of $\mathcal{S}$ is equal to the $n$th element of $\mathbf{E}_a(\{1,1,\ldots,1\})$ if the non-zero element in the $n$th row of $E^{\cos\theta}$ is a `1', and otherwise is equal to the $n$th element of $\mathbf{E}_b(\{1,1,\ldots,1\}$.  The following properties of the defined `addition' operation are easy to show:

\begin{itemize}
\item 
When $\theta=0$, $\mathcal{S}=\mathbf{E}_a(\{1,1,\ldots,1\})$ ;when $\theta=\pi/2$, $\mathcal{S}=\mathbf{E}_b(\{1,1,\ldots,1\})$ 
\item
With $\cos\theta \in \mathbbm{Q}_2$, the coefficient of correlation between $\mathcal{S}$ and $\mathbf{E}_a(\{1,1,\ldots,1\})$ is equal to $\cos\theta$. With $\theta$ varying uniformly in time, the phenomenon of wave interference is manifest. 
\item
If $0<\cos\theta < \pi/2 \in \mathbbm{Q}_2$, then $\sin\theta \notin \mathbbm{Q}_2$ and vice versa; see Appendix A.
\item
If $\sin\theta \in \mathbbm{Q}_2$ then $\mathcal{S}$ is as (\ref{eq:superp}), with $\mathbf{E}_a$ swapped with $\mathbf{E}_b$.
\item
The definition is distributive in the sense that  
\be
\mathbf{E}^{\alpha}(\mathcal{S})=(\cos\theta\  \mathbf{E}^{\alpha}\mathbf{E}_a + \sin\theta\  \mathbf{E}^{\alpha}\mathbf{E}_b) \{1,1,\ldots,1\}
\ee
\end{itemize}

On this basis, we can write, for example
\begin{eqnarray}
|\uparrow\rangle+e^{i\theta}|\downarrow\rangle &\mapsto& \mathbf{E}_a (\cos\theta\ \mathbf{E}_0+\sin\theta\ \mathbf{E}\mathbf{E}_0)(\{1,1,\ldots,1\})\nonumber\\
&=& (\cos\theta\ \mathbf{E}_a + \sin\theta\ \mathbf{E}_a\mathbf{E}) (\{1,1,\ldots,1\})\nonumber\\
&=& (\cos\theta\ \mathbf{E}_a + \sin\theta\ \mathbf{E}_b) (\{1,1,\ldots,1\})=\mathcal{S}
\end{eqnarray}
from (\ref{eq:superp}).

A key point in this proposed reformulation of Hilbert space, is that the notion of `adding wavefunctions' does not imply `superposition' of states, with all the (Schr\"{o}dinger's cat) paradoxes that that implies. Rather, the $n$th element (beable) of the bit string `$\mathcal{A}+i\mathcal{B}$' is formed from the $n$th elements of the bit strings $\mathcal{A}$ and $\mathcal{B}$, taking account of the hyper-complex structure of the associated permutation operators $\mathbf{E}_j$. Because of this complex structure, the $n$th elements of $\mathcal{A}+i\mathcal{B}$ need equal neither the $n$th elements of $\mathcal{A}$ nor $\mathcal{B}$. Note that no additional collapse hypothesis is required to obtain definite beable elements.  

\subsubsection{2 Qubits}
   
In quantum theory, the general 2-qubit state is given by
\be
\label{eq:2qubit}
|\psi\rangle=p_0|\uparrow\uparrow\rangle+p_1|\uparrow\downarrow\rangle +p_2|\downarrow\uparrow\rangle+p_3|\downarrow\downarrow\rangle
\ee 
where $p_i \in \mathbbm{C}$ satisfy $|p_0|^2+|p_1|^2 +|p_2|^2+|p_3|^2=1$, giving a Hilbert space with seven degrees of freedom, including the global phase degree of freedom, modulo which gives the standard complex-three dimensional projective Hilbert space $\mathbbm{CP}_3$. In the reformulation proposed here, the wavefunction of a 2-qubit state (in a universe of $N$ qubits) is given by two $2^N$-long bit strings $\mathcal{S}_1$ and $\mathcal{S}_2$. Here the seven degrees of freedom are represented by recalling from (\ref{eq:m3}) that any ${E}_j \ j>1$ belongs to a set of seven square roots of minus one whose elements can be listed as $\{E, E_A, E_B, E_C, E_D, E_E, E_F\}$. As before we use roots of $E$ to represent global $`U(1)'$ invariance. Consistent with this, the coefficient of correlation between $\mathcal{S}_1$ and $\mathcal{S}_2$ is invariant under the global transformation
\be
\label{eq:zphase}
\mathcal{S}_1 \mapsto \mathbf{E}^{\alpha}(\mathcal{S}_1), \   \mathcal{S}_2 \mapsto \mathbf{E}^{\alpha}(\mathcal{S}_2)
\ee
where $\alpha \in \mathbbm{Q}_2$. The remaining six degrees of freedom can be described by considering bit strings which combine (in the sense defined by (\ref{eq:superp})), the seven `basis' sequences
\begin{eqnarray}
\{1,1,&\ldots&,1\},\nonumber\\
\mathbf{E}_A(\{1,1,\ldots,1\} ), \mathbf{E}_B(\{1,1,&\ldots&,1\} ), \mathbf{E}_C(\{1,1,\ldots,1\} ),\nonumber\\
\mathbf{E}_D(\{1,1,\ldots,1\} ), \mathbf{E}_E(\{1,1,&\ldots&,1\} ), \mathbf{E}_F(\{1,1,\ldots,1\} )
\end{eqnarray}
Representing the wavefunction $|\psi\rangle$ for two qubits as the pair $\{\mathcal{S}_1, \mathcal{S}_2\}$, the qubits will be said to be entangled if $\mathcal{S}_1$ is correlated with $\mathcal{S}_2$. Consider, for example
\begin{eqnarray}
\label{eq:entangled}
\mathcal{S}_1&=&\mathbf{E}_A(\{1,1,\ldots,1\})\nonumber \\
\mathcal{S}_2&=&(\cos\theta\  \mathbf{E}_A + \sin\theta\  \mathbf{E}_D) \{1,1,\ldots,1\}
\end{eqnarray}
By the discussion in section \ref{sec:1qubit}, if $\cos\theta \in \mathbbm{Q}_2$ then the correlation between $\mathcal{S}_1$ and $\mathcal{S}_2$ is equal to $\cos\theta$. 

\subsubsection{$N$ Qubits}

Continuing to larger $N$, the wavefunction of an entire universe of $N$ qubits is represented by a set $\{\mathcal{S}_1, \mathcal{S}_2, \ldots, \mathcal{S}_N\}$ of bit strings each of length $2^N$ - equivalently, as discussed above, as a rational $\mathcal{R}_U$ with $2^N$ base-$2^N$ digits. In standard quantum theory, the Hilbert space of $N$ qubits has dimension $2^N-1$, including one global phase degree of freedom. In the reformulation, we have the set $\{E, E_2 \ldots E_{2^N-1}\}$ of $2^N-1$ square roots on minus one. As before, the the global phase degree of freedom is represented by the roots $E^{\alpha}$ of $E$. The remaining degrees of freedom are associated with linear combinations of the basis sequences

\begin{eqnarray}
\{1,1,&\ldots&,1\},\nonumber\\
\mathbf{E}_2(\{1,1,\ldots,1\} ), \mathbf{E}_3(\{1,1,&\ldots&,1\} ), \ldots \mathbf{E}_{2^N-1}(\{1,1,\ldots,1\} )
\end{eqnarray}
as in (\ref{eq:superp}). 

It can be asked whether the proposed reformulation is testably different from quantum theory. One interesting fact, which may be relevant in this respect, emerges at the level of $4$ qubits. Unlike the state space of $1$, $2$ or $3$ qubits, the Hilbert space $\mathbbm{S}^{31}$ of 4 qubits in standard quantum theory cannot be Hopf fibrated, due to a theorem of Adams and Atiyah (1966). As Bernevig and Chen (2003) note, the failure of the Hilbert space to fibrate appears to lead to fundamental difficulties in describing the entanglement structure of 4 or more qubits. By contrast, the present theory is constrained by neither the continuum properties of hyper-complex algebraic fields, nor their corresponding topological spaces. It is interesting to note that $4$-qubit structures are needed to describe the quantum of gravity. This indicates that the proposed reformulation may more readily incorporate the effects of gravity than does does conventional quantum theory. 

\subsubsection{Relation to the Idealised Model of Quantum Measurement}

The construction of $S_0$ and $S_{\theta}$ in section \ref{sec:toy} is an idealisation of the hyper-complex permutation operators developed here. A more precise linkage bewteen $S_0$ and $S_{\theta}$ and the proposed reformulation of quantum theory, can be given as follows. In terms of the proposed 2-qubit reformulation of Hilbert space put $|\psi\rangle \mapsto \{\mathcal{S}, \mathcal{S}_{\theta}\}$, where
\begin{eqnarray}
\mathcal{S}&=&\mathbf{E}_a\{1,1,\ldots,1\}\nonumber\\
\mathcal{S}_{\theta}&=&(\cos \theta \ \mathbf{E}_a+\sin\theta \ \mathbf{E}_b)\{1,1,\ldots,1\},
\end{eqnarray}
where $\{E, E_a, E_b\}$ denotes a pure imaginary quaternionic triple in the space of $2^N-1$ hyper-complex permutation operators. 
 
Using (\ref{eq:root}), define
\begin{eqnarray}
\mathbf{E}^{\alpha}(\mathcal{S})&=&\{a_1, a_2,\ldots, a_{2^N}\}\nonumber\\
\mathbf{E}^{\alpha}(\mathcal{S}_{\theta})&=&\{a_1(\theta), a_2(\theta), \ldots, a_{2^N}(\theta)\}
\end{eqnarray}
and put 
\begin{eqnarray}
S_0(\frac{\alpha\pi}{2})&=&a_1 \nonumber\\
S_{\theta}(\frac{\alpha\pi}{2})&=&a_1(\theta)
\end{eqnarray}
that is, $S_{\theta}(\alpha\pi/2)$ denotes the leading bit of $\mathbf{E}^{\alpha}(\mathcal{S}_{\theta})$. 

When $\theta=0$, then $S_0$ and $S_{\theta}$ are identical. When $\theta=\pi/2$ then $\mathcal{S}_{\theta}=\mathbf{E}_b\{1,1,\ldots,1\}=\mathbf{E}\mathbf{E}_a\{1,1,\ldots,1\}=\mathbf{E}(\mathcal{S})$. Hence, $S_{\pi/2}(\alpha\pi/2)=S_0(\pi/2+\alpha\pi/2)$. When $\theta=\pi$, then $\mathcal{S}_{\theta}=-\mathbf{E}_a\{1,1,\ldots, 1\}=\mathbf{E}^2(\mathcal{S})$, hence $S_{\pi}(\alpha\pi/2)=S_0(\pi+\alpha\pi/2)$. When $\theta=3\pi/2$ then $\mathcal{S}_{\theta}=-\mathbf{E}_b\{1,1,\ldots,1\}=\mathbf{E}^3(\mathcal{S})$. Hence, $S_{3\pi/2}(\alpha\pi/2)=S_0(3\pi/2+\alpha\pi/2)$. 

For all other values of $\theta$, $\mathcal{S}_{\theta}$ is never equal to $\mathbf{E}^{\alpha}(\mathcal{S})$ for any $\alpha$ - since $\mathbf{E}^{\alpha}$ induces a cyclic displacement of the elements of $\mathcal{S}$, there is no $\alpha$ where $\mathbf{E}^{\alpha}(\mathcal{S})$ is partially correlated with $\mathcal{S}$. That is, the values of $\theta$ for which $\mathcal{S}_{\theta}$ belongs to the qubit equivalence class $\{\mathbf{E}^{\alpha}(\mathcal{S}):\  \alpha \in \mathbbm{Q}_2\}$, are precisely the values $\theta$ for which $\theta$ is a dyadic rational multiple of $\pi$ and $\cos\theta$ is dyadic rational ie $\{0,\pi/2, \pi, 3\pi/2\}$. This result links the idealised model in section \ref{sec:toy} with the proposed reformulation of quantum theory, and makes the result on causal incompleteness relevant to the interpretation of Bell's theorem in (this reformulation of) quantum theory.  

\section{Incompleteness and the Metaphysics of Free Will}
\label{sec:meta}

It has been proposed that inherent mathematical incompleteness (of the type describing the sets of rational numbers, Turing-computable functions, arithmetic proofs and so on) provides a new interpretation of the experimental violation of the Bell inequalities, one that does not invoke or require non-local causality. This interpretation applies to a class of physical theory for which state space is inherently granular and discontinuous. Within such class of theory, the freedom of experimenters to choose measurement settings is tempered by the granular structure of state space. The proposed interpretation of the Bell inequalities is that they reveal the inevitable mathematical incompleteness of the causal structure underlying such theory. The kinematic structure of quantum theory has been reformulated so that it then belongs to this class of theory. 

 It is well known that violation of the Bell inequalities can be `explained' if the notion of free will is completely rejected. This is generally considered an unsatisfactory explanation, for reasons which go under the general description `conspiratorial'. In an attempt to clarify this issue, Bell (1993) considers a deterministic dynamical system which replaces the whimsical experimenter. This system selects between two possible outputs $a$ or $a'$ on the basis of the parity of the digit in the millionth decimal place of some input variable. Then, fixing $a$ or $a'$ fixes something about the input - ie whether the millionth digit is odd or even. Bell's objection to a deterministic explanation of the violation of his eponymous inequalities is this: this peculiar piece of information, the millionth digit, is unlikely to be the vital piece of information for any distinctively different purpose ie it is otherwise rather useless. 

However, in the reformulation of quantum kinematics, the wavefunction of the universe is constructed from finite $N$ bit strings each of length $2^N$, and the equivalent of unitary transformations involve permutation and related operators acting on these bit strings. Typically these permutation operators, represented as matrices, have terms on the anti-diagonal. For example, the global phase operator $\mathbf{E}^{1/2^N}$, see (\ref{eq:root}), acting on some sequence $\mathcal{S}$ brings to the front of the sequence, an element that was previously at the back of the sequence. That is, at the heart of our reformulation of the complex Hibert space of quantum theory, are operators whose action is very similar, in essence, to Bell's deterministic dynamical system. The entanglement structure of the universe is given by precise intricate relationships between the bits of the different bit strings. In this perspective, bits near the end of a bit string are no less `vital' for `distinctively different purposes' than bits near the front of the bit string. Like a Sudoku puzzle, violate one piece of the structure (either at the beginning or end of a bit string) and you violate the structure everywhere. 

Rather, the real difficulty with deterministic explanations of the violation of the Bell inequalities is their contradiction of our strong intuition that the experiment could have been performed differently. In Bell's example above, our intuition suggests the input could have been otherwise, at least as far as the trailing digits of the input number are concerned. Any explanation of the violation of the Bell inequality which does not address this deeply-held feeling, is not likely to be accepted. 

In the current proposal, our intuition about free will is not rejected \emph{per se}, but rather (as far as the EPR-Bohm-Bell experiments are concerned) is derived from the computational properties of the models $Sp_1$ and $Sp_2$, see (\ref{eq:cf}). Hence, our intuition infers that the experimenter could have chosen from an arbitarily dense set of alternative detector orientations to the one actually chosen, and, from the properties of $Sp_1$, this belief is not inconsistent with the (proposed reformulation of the) laws of physics. Since $Sp_1(0, \lambda)$ defines `reality', $Sp_1(\pi, \lambda)$ defines an alternate world precisely anitcorrelated with reality, and $Sp_1(\pi/2, \lambda)$ and $Sp_1(3\pi/2, \lambda)$ define alternate worlds uncorrelated with reality. However, if $Sp_1(\delta\theta_1, \lambda)$ solved algorithmically for a perturbed orientation such as required to derive a Bell inequality, $Sp_1$ would never halt. That is, in circumstances where our intuition might contradict physics, our intuition, acting computationally, would never be able to ascertain what the relevant measurement outcome would have been. 

On the other hand, our cognitive reasoning is (from time to time, at least) able to transcend such a purely computational perspective. Does an awareness of algorithmic incompleteness imply that non-computability is a feature of such cognitive reasoning, and by implication a feature of the laws of physics, as has been suggested by Penrose (1994)? In fact, it would be hard to reconcile this notion with this paper's underlying premise that the granular reformulation of quantum theory is ultimately finite (with the wavefunction of the universe being given by a sequence of $2^N$ base-$2^N$ digits, for some very large but nevertheless finite $N$). A possible alternative suggestion, therefore, is that an awareness of algorithmic incompleteness may instead arise from some cognitive ability to jump (perhaps involuntarily) between computationally-inequivalent finite systems. For example, in the model developed here, a finite division of the circle based on angular segments which are equal dyadic rational fractions of $\pi$, is not equivalent to a finite division of the cirle based on angular segments whose cosines are equal dyadic rational fractions ie based on equal divisions of the diameter. From an awareness of both finite cyclotomies one can recognise the inability of one to contain the other. 

In conclusion, it is ironic, perhaps, that exploitation of mathematical incompleteness in physical theory may turn out to be the key notion that allows EPR's goal to be achieved, of developing a theory of the quantum that is more physically complete than standard quantum theory.  

\appendix
\section{A fundamental property of the cosine function}

The discussion in section \ref{sec:toy} uses a rather basic property of the cosine function, albeit one rarely (if ever?) used in physics.  For completeness, we give a simple proof of this property, based on unpublished work by Jahnel (2004). It can be seen as a special example from the theory of trigonometric diophantine equations (Conway and Jones, 1976).  

\textbf{Theorem} Let $0<\theta/\pi<1/2 \in \mathbbm{Q}_2$, then $\cos \theta \notin \mathbbm{Q}_2$. 

With $0<\theta/\pi<1/2 \in \mathbbm{Q}_2$, assume that $2\cos \theta =a/b$ is rational, where $a,b \in \mathbbm{Z}, b \ne 0$ have no common factors. 

Using the identity
\be
2\cos 2\theta = (2\cos\theta)^2-2
\ee
we have
\be
2\cos 2\theta=\frac{a^2-2b^2}{b^2}
\ee
Now $a^2-2b^2$ and $b^2$ have no common factors, since if $p$ were a prime number dividing both, $p|b^2 \Rightarrow p|b$ and $p|(a^2-2b^2) \Rightarrow p|a$, a contradiction. 

Hence, if $b \ne \pm1$, then the denominators in $2\cos\theta$, $2\cos2\theta$, $2\cos4\theta$, $2\cos8\theta\ldots$ get bigger and bigger without limit. On the other hand, $\theta/\pi=m/n$ which means that the sequence $(2\cos 2^k \theta)_{k\in \mathbbm{N}}$ admits at most $n$ values. Hence we have contradiction. Hence $b=\pm1$. Hence $\cos\theta=0, \pm1/2, \pm1$. No $0<\theta/\pi<1/2 \in \mathbbm{Q}_2$ has $\cos\theta$ with these values. QED.

Finally note that Pythagorean integer triples $\{x,y,z\}$ satisfying $x^2+y^2=z^2$ can be parametrised as $x=2uv, y=u^2-v^2, z=u^2+v^2$ where $(u, v)$ are integers without common factor and of different parity (eg Hardy and Wright, 1979). Hence, if $0 < \theta < \pi/2$ and both $\cos \theta$ and $\sin \theta$ are rational, then $\cos \theta = 2uv/(u^2+v^2)$ and $\sin\theta=(u^2-v^2)/(u^2+v^2)$. Since $u$ and $v$ are of different parity, then $u^2+v^2$ cannot be divided by 2. Hence $\cos \theta$ and $\sin\theta$ cannot both be dyadic rational.    

\section*{References}

\begin{description}

\item Adams, F.J. and Atiyah, M.F.A. 1966: On K-theory and Hopf invariant. Quarterly J. Math., 17, 31-8.  

\item Bernevig, B.A. and Chen H.-D., 2003: Geometry of the three-qubit state, entanglement and division algebras. J.Phys A, 36, 8325-8339.

\item Bell, J.S., 1993: Free variables and local causality. In `Speakable and unspeakable in quantum mechanics.'
Cambridge University Press. 212pp.

\item Bohm, D., 1980: Wholeness and the implicate order. Routledge and Kegan Paul, London. 

\item Conway, J.H. and Jones, A.J., 1976: Trigonometric Diophantine Equations. Acta Arithmetica, 30, 229-240.

\item Hardy, G.H. and Wright, E.M., 1979: The Theory of Numbers. Oxford University Press.

\item Jahnel, J., 2005: When does the (co)-sine of a rational angle give a rational number? Available online at www.uni-math.gwdg.de/jahnel/linkstopaperse.html

\item Kane, R., 2002: The Oxford Handbook on Free Will. Oxford University Press. 638pp

\item Menzies, P. 2001: Counterfactual Theories of Causation. The Stanford Encyclopedia of Philosophy (Spring 2001 Edition), Edward N. Zalta (ed.), URL = http://plato.stanford.edu/archives/spr2001/entries/causation-counterfactual/

\item Palmer, T.N., 2004: A granular permutation-based representation of complex numbers and quanternions: elements of a possible realistic quantum theory. Proc. Roy. Soc., 60A, 1039-1055.

\item Penrose, R., 1994: Shadows of the mind. Oxford University
Press. Oxford. 450pp

\item Price, H., 1996: Time's Arrow and Archimedes Point. Oxford University Press. 306pp

\item Rae, A.I.M., 1992: Quantum Mechanics. Institute of Physics. Bristol. 

\item Wheeler, J.A., 1994: In `Physical Origins of Time Asymmetry, ed J.J. Halliwell, J.Peres-Mercader and W.H. Zurek, pp. 1-29. Cambridge University Press. 

\end{description}
\end{document}